\documentclass[vecphys]{svmult}

\usepackage{makeidx}         
\usepackage{graphicx}        
\usepackage{multicol}        
\usepackage[bottom]{footmisc}

\makeindex             

\def\farcs{\hbox{$.\!\!^{\prime\prime}$}}
\DeclareMathAlphabet{\mathsc}{OT1}{cmr}{m}{sc}
\def\testbx{bx}%
\DeclareRobustCommand{\ion}[2]{%
\relax\ifmmode
\ifx\testbx\f@series
{\mathbf{#1\,\mathsc{#2}}}\else
{\mathrm{#1\,\mathsc{#2}}}\fi
\else\textup{#1\,{\mdseries\textsc{#2}}}%
\fi}
\def\arcsec{\hbox{$^{\prime\prime}$}}

\begin{document}
\title*{A Look into the Guts of Sunspots}
\author{L.R.\ Bellot Rubio}
\institute{Instituto de Astrof\'{\i}sica de Andaluc\'{\i}a (CSIC), 
Apdo.\ 3004, 18080 Granada, Spain
\texttt{lbellot@iaa.es}}
\maketitle
\index{L.R.\ Bellot Rubio}

\vspace*{-7cm}
\noindent
{\small To appear in {\em Highlights of Spanish Astrophysics IV}, 2007, 
eds.\ F.\ Figueras, J.M.\ Girart, M.\ Hernanz, C.\ Jordi (Springer)}

\vspace*{6.2cm} Advances in instrumentation have made it possible to study
sunspots with unprecedented detail. New capabilities include imaging
observations at a resolution of 0\farcs1 (70~km on the sun), spectroscopy at
$\sim 0\farcs2$, and simultaneous spectro\-polarimetry in visible and infrared
lines at resolutions well below~1\arcsec.  In spite of these advances, we
still have not identified the building blocks of the penumbra and the
mechanism responsible for the Evershed flow. Three different models have been
proposed to explain the corpus of observations gathered over the years.  The
strengths and limitations of these models are reviewed in this contribution.

\section{Introduction}
\begin{figure}[t]
\centering

\includegraphics[height=6.1cm,bb= 55 35 330 325,clip]{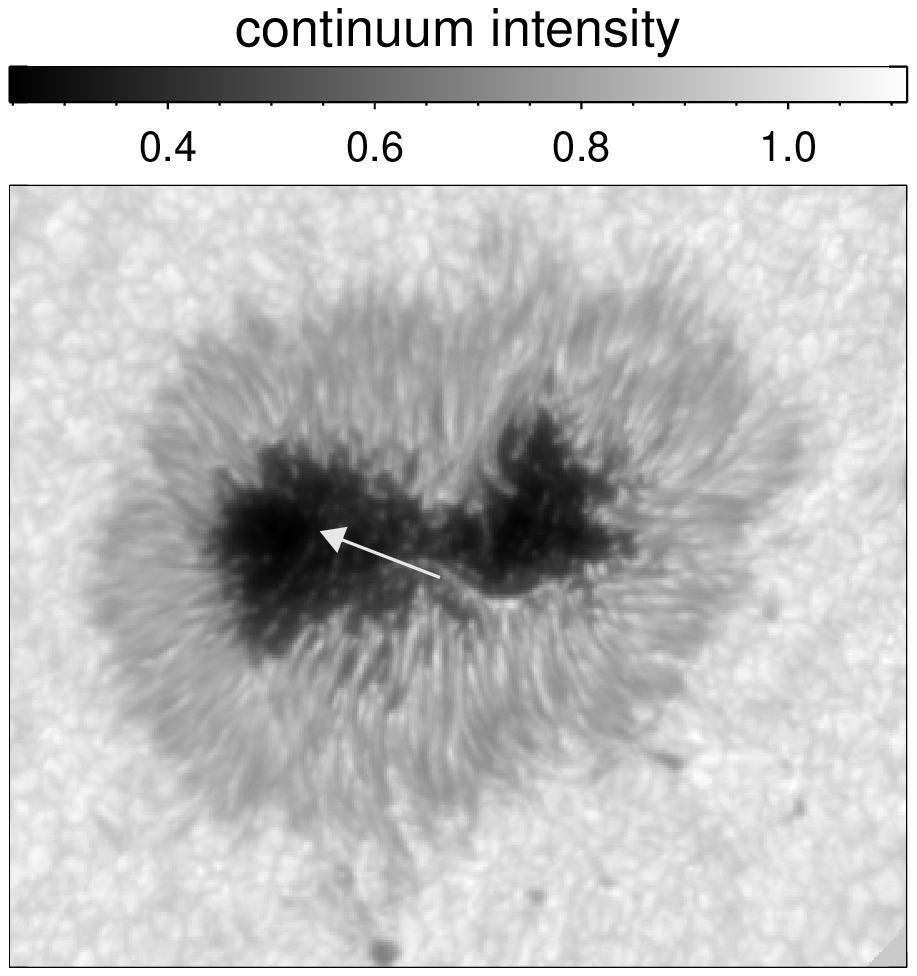}
\includegraphics[height=6.1cm,bb= 55 35 330 325,clip]{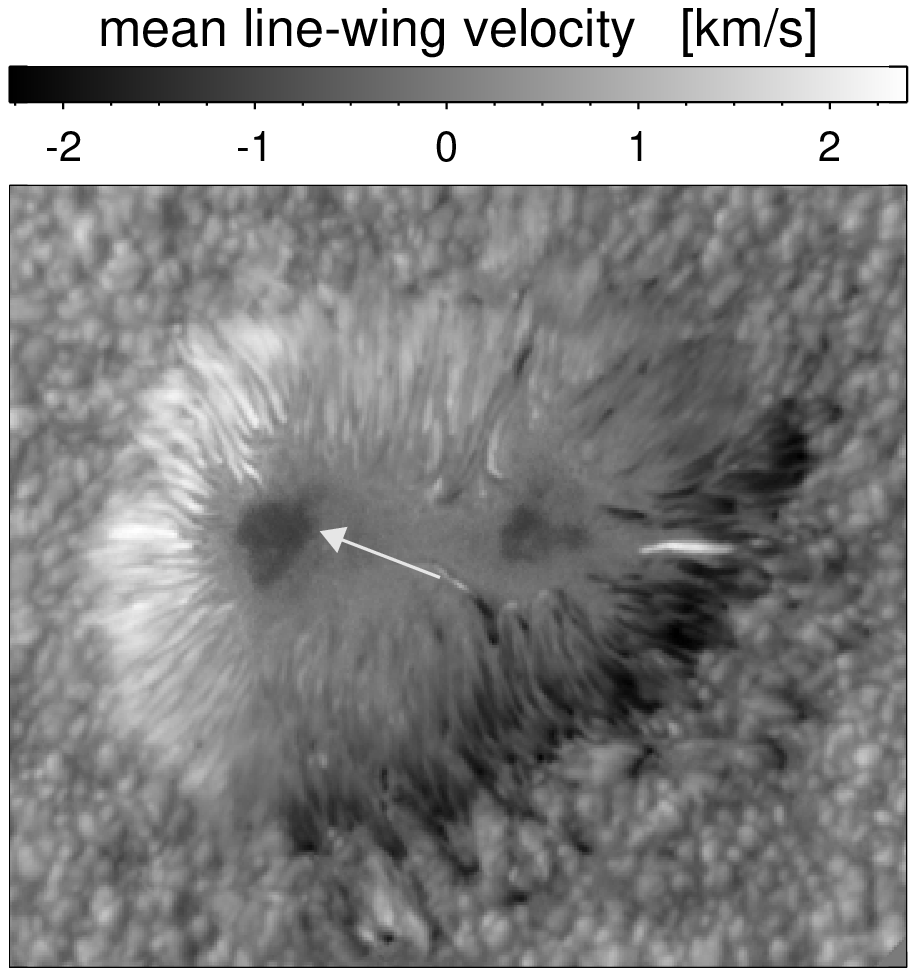}
\caption{AR 10905 as observed with IBIS at the DST of NSO/Sac Peak Observatory
on Aug 24, 2006. The spot was located $42^{\rm o}$ off the disk center. The
spatial resolution is about 0\farcs3. The observations were taken in the
\ion{Fe}{i} 709.0~nm line. {\em Left:} Continuum image.  {\em Right:}
Dopplergram derived from line-wing intensities. 
Positive velocities indicate blueshifts. The arrow points to disk
center. Blue\-shifts in the center-side and redshifts in the limb-side
penumbra are the signatures of the Evershed flow. Observations and data
reduction courtesy of A.\ Tritschler and H.\ Uitenbroek.}
\label{ibis}  
\end{figure}

Sunspots were the first celestial objects known to harbor magnetic fields, a
discovery made by Hale in 1908 \cite{hale}. One year later, Evershed described
a nearly horizontal plasma outflow in sunspot penumbrae \cite{evershed}. This
flow produ\-ces the so-called Evershed effect: redshifted spectral lines in the
limb-side penumbra and blueshifts in the center-side penumbra
(Fig.~\ref{ibis}). As seen in conti\-nuum images, the penumbra is formed by
bright and dark filaments oriented radially. Observations have revealed a 
close relationship between the filamentary structure of the penumbra, its
magnetic field, and the Evershed flow.

The penumbra exhibits a complex magnetic topology, with fields of different
strengths and inclinations interlaced both vertically and horizontally (see
\cite{solanki} and \cite{bellotbiermann} for reviews). The more inclined
fields channel the Evershed flow, while the more vertical fields are not
associated with significant mass motions.  In the inner penumbra, the magnetic
field and the flow are directed upward~\cite{rimmele,schliche+schmidt,
bellotetal2004,bellotetal2005b,rimmelemarino}, but in the outer penumbra one
observes downward flows~\cite{rimmele,schliche+schmidt,schlicheetal2004} along
magnetic field lines returning back to the solar
surface~\cite{westend,mathewetal2003, bellotetal2004, borreroetal2005,
langhansetal2005,monica}.  The vertical interlacing of different magnetic
field components with different velocities is responsible for the non-zero net
circular polarization (NCP) of spectral lines emerging from the
penumbra. 

These ingredients led to the concept of {\em uncombed penumbra}
\cite{solankimontavon} (see also~\cite{title} and \cite{valentin}). Basically,
an uncombed penumbra consists of nearly horizontal magnetic flux tubes
embedded in a stronger and more vertical ambient field. The tubes carry the
Evershed flow, with the ambient field being essentially at rest.  The uncombed
penumbral model is supported by numerical simulations of interchange
convection (\cite{schliche} and references therein), but the detection of
individual flux tubes in spectropolarimetric observations has proven elusive
due to their small sizes (100--200~km in diameter).

\begin{figure}[t]
\centering
\includegraphics[height=5.2cm]{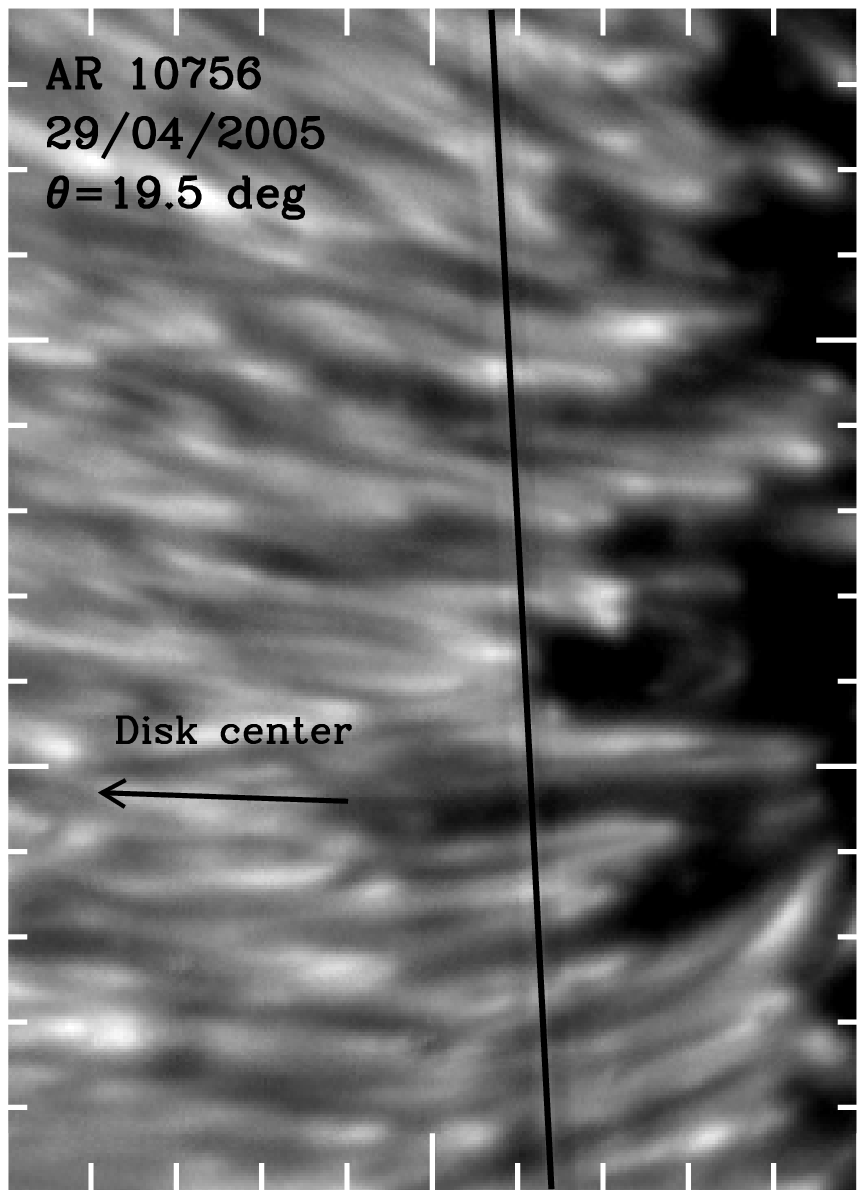}
\includegraphics[height=5.2cm]{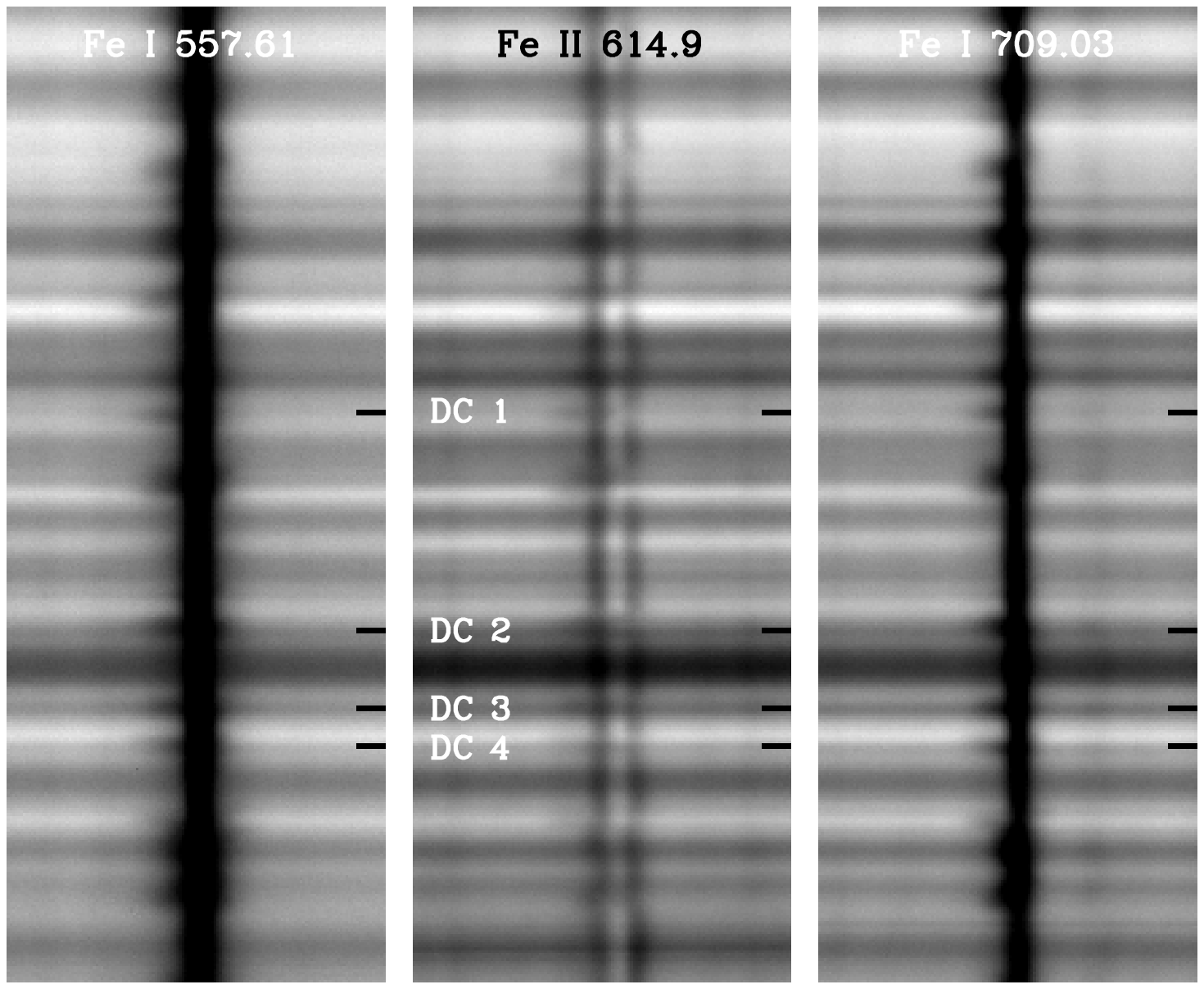}
\caption{ Multi-line spectroscopy of dark-cored penumbral filaments at
0\farcs2 resolution. The data were taken at the SST on April 29, 2005, and
correspond to the center-side penumbra of AR 10756. {\em Left:} Slit-jaw
image.  The slit crosses four dark-cored filaments. {\em Right:} Intensity
profiles of \ion{Fe}{i} 557.6, \ion{Fe}{ii} 614.9, and \ion{Fe}{i} 709.0~nm
along the slit. The dark cores (``DC'') are marked with small horizontal 
lines. Their large blueshifts are produced by Evershed flows directed 
upward. See~\cite{bellotetal2005} for details.}
\label{spectra}  
\end{figure}

Recently, high-resolution (0\farcs1--0\farcs2) images taken with the Swedish
1-m Solar Telescope and the Dutch Open Telescope on La Palma have demonstrated
that many penumbral filaments possess internal structure in the form of a dark
core \cite{scharmer,sutterlin}. The dark core is surrounded by two narrow
lateral brightenings (Fig.~\ref{spectra}, left), both of which are observed to
move with the same speed and direction as a single entity.  The fact that the
various parts of dark-cored filaments show a coherent behavior have raised
strong expectations that they could be the fundamental constituents of the
penumbra, i.e., the flux tubes postulated by the uncombed model. Spectroscopy
at 0\farcs2 resolution suggests that the Evershed flow is stronger in the dark
cores (Fig.~\ref{spectra}, right) and that dark-cored filaments possess weaker
fields than their surroundings close to the umbra~\cite{bellotetal2005}. 
Other than that, the magnetic and kinematic properties of dark-cored penumbral
filaments remain unknown, so for the moment it is not possible to confirm or
reject the idea that they represent individual tubes.

In the meantime, alternative models of the penumbra have emerged: scenarios
based on MIcro-Structured Magnetic Atmospheres (MISMAs; \cite{jorge98,jorge}) and
field-free gaps (the gappy penumbral model; \cite{spruit+scharmer}). These
models try to explain the morphological and spectropolarimetric properties of the
penumbra. They also claim to solve important problems of the uncombed model. In
the following, the strengths and limitations of the different models are
examined.

\section{Competing penumbral models}
\subsection{Uncombed model}
\begin{figure}[t]
\centering
\includegraphics[height=3.18cm]{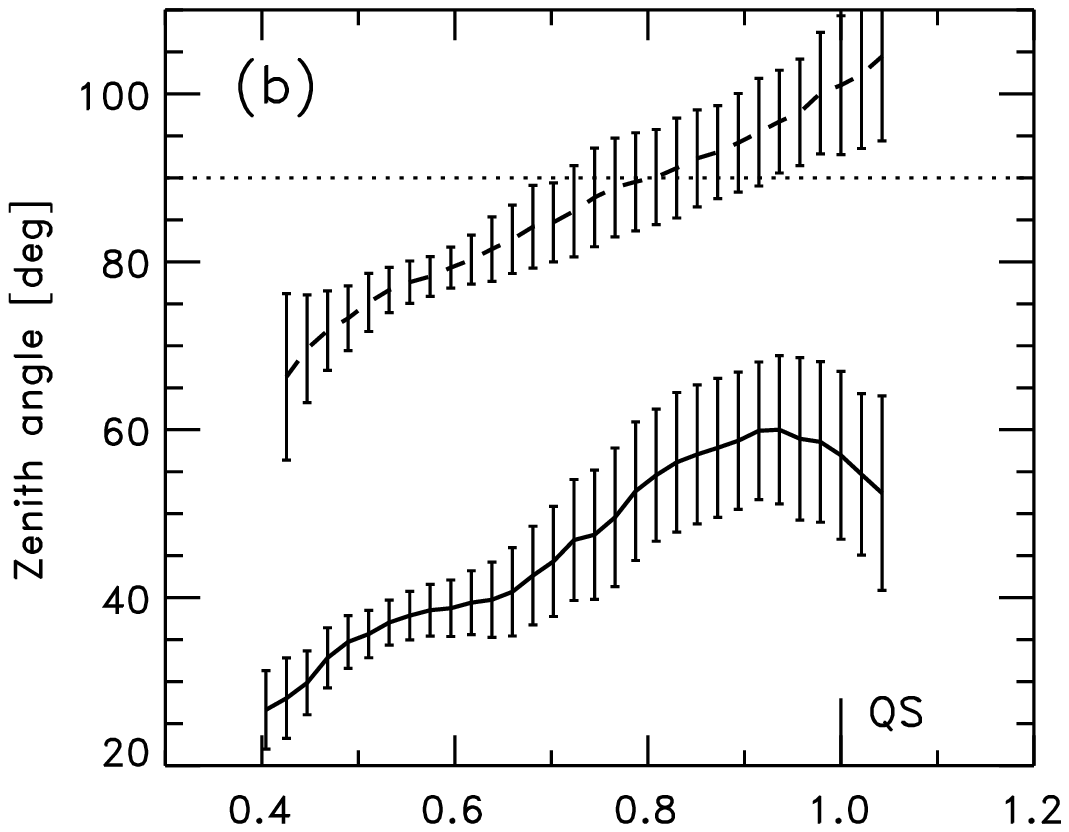}
\includegraphics[height=3.18cm]{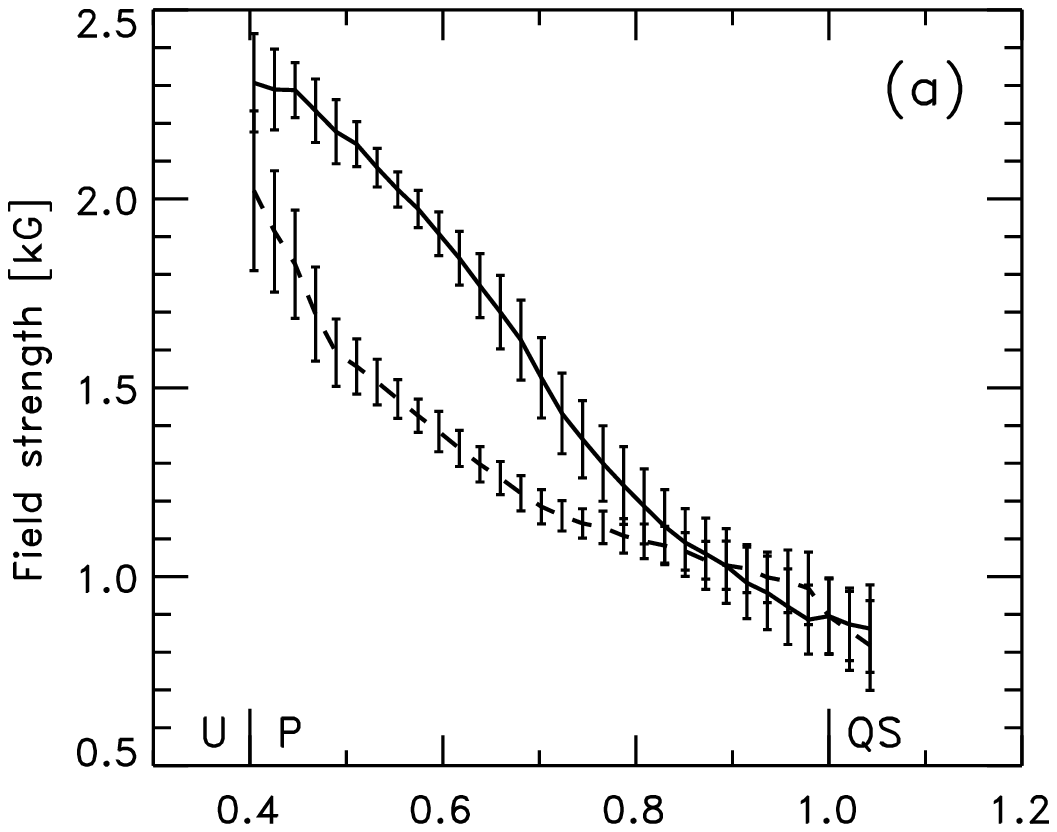}
\includegraphics[height=3.18cm]{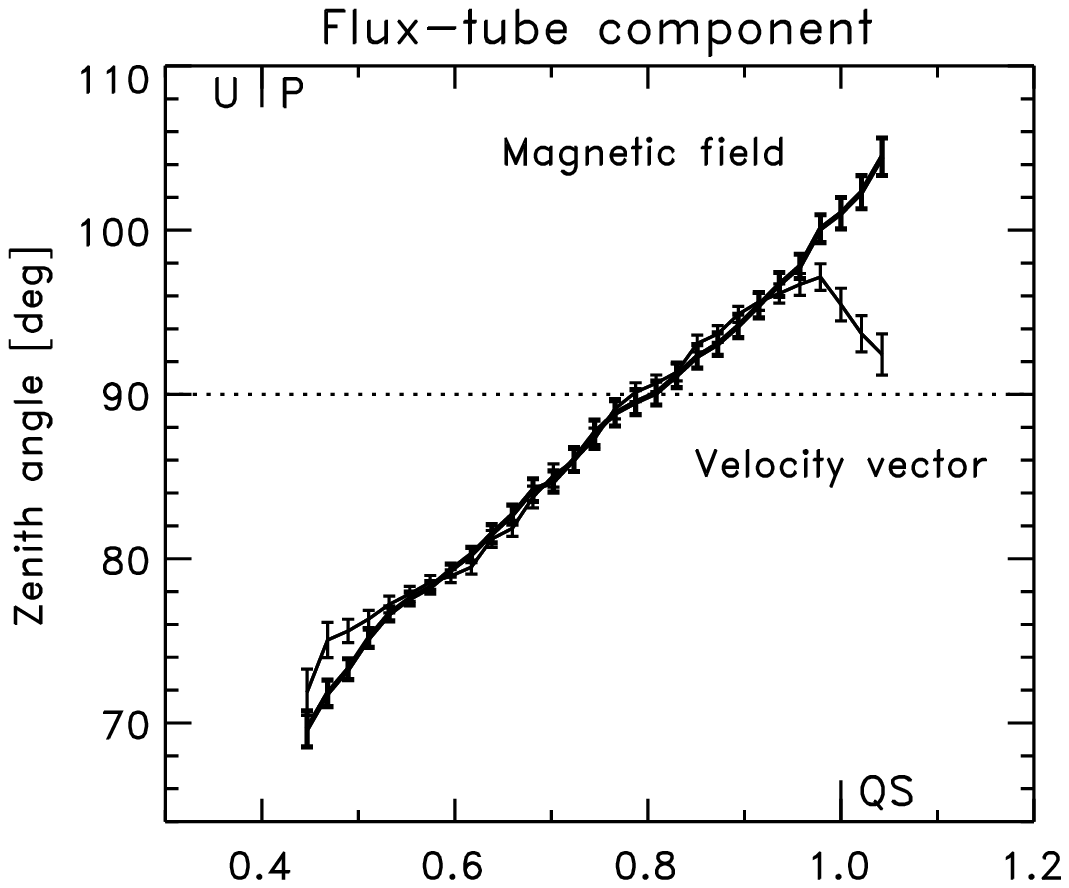}
\caption{Radial variation of the field inclination ({\em left}) and field
strength ({\em center}) in the penumbra of AR 8704 as derived from a
two-component inversion of the \ion{Fe}{i} lines at 1565~nm. Solid and dashed
lines represent the flux-tube and ambient atmospheres. {\em Right:}
Inclination of the velocity vector in the flux-tube component. Taken from
\cite{bellotetal2004}.}
\label{structure}  
\end{figure}

\begin{figure}[t]
\centering
\includegraphics[height=2.55cm,bb=70 25 473 273]{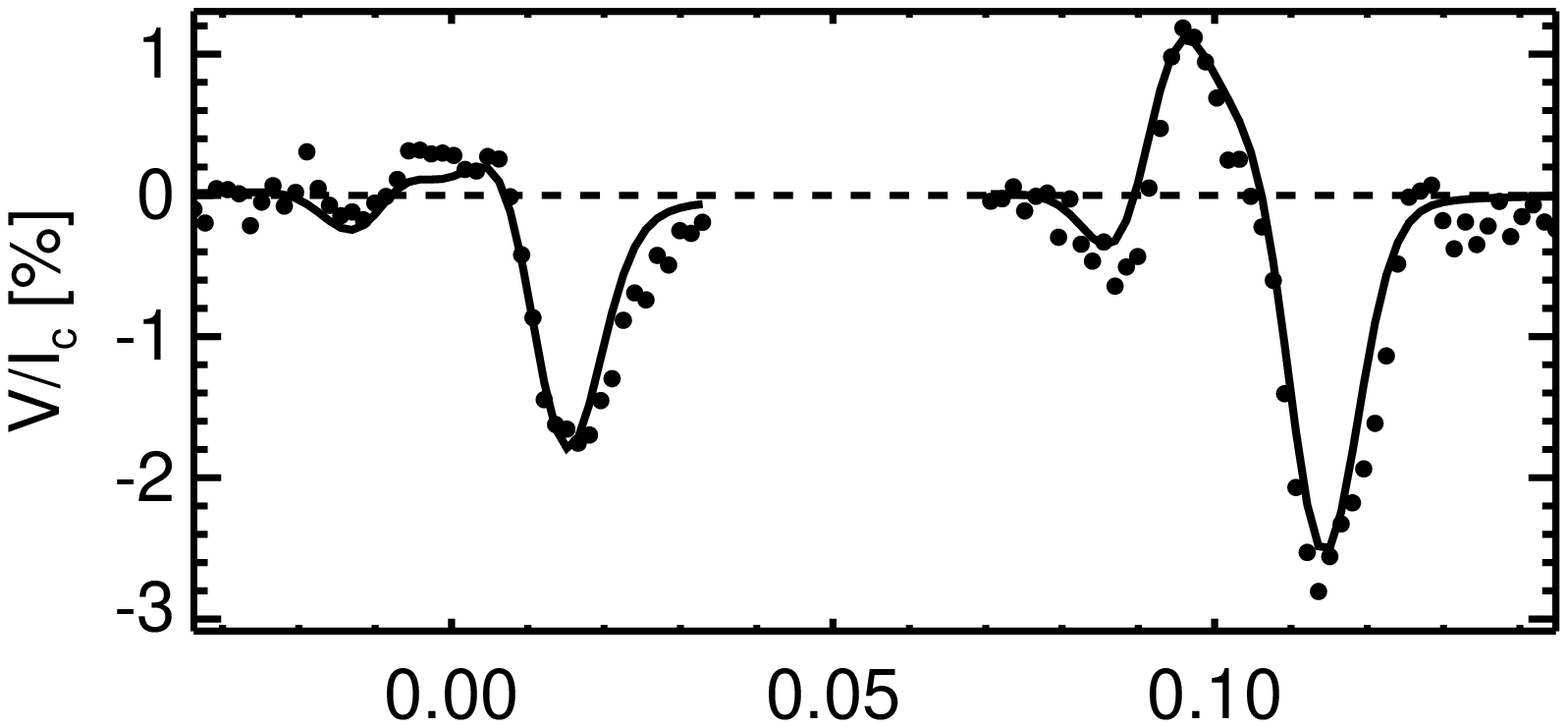}
\includegraphics[height=2.55cm,bb=30 25 508 273]{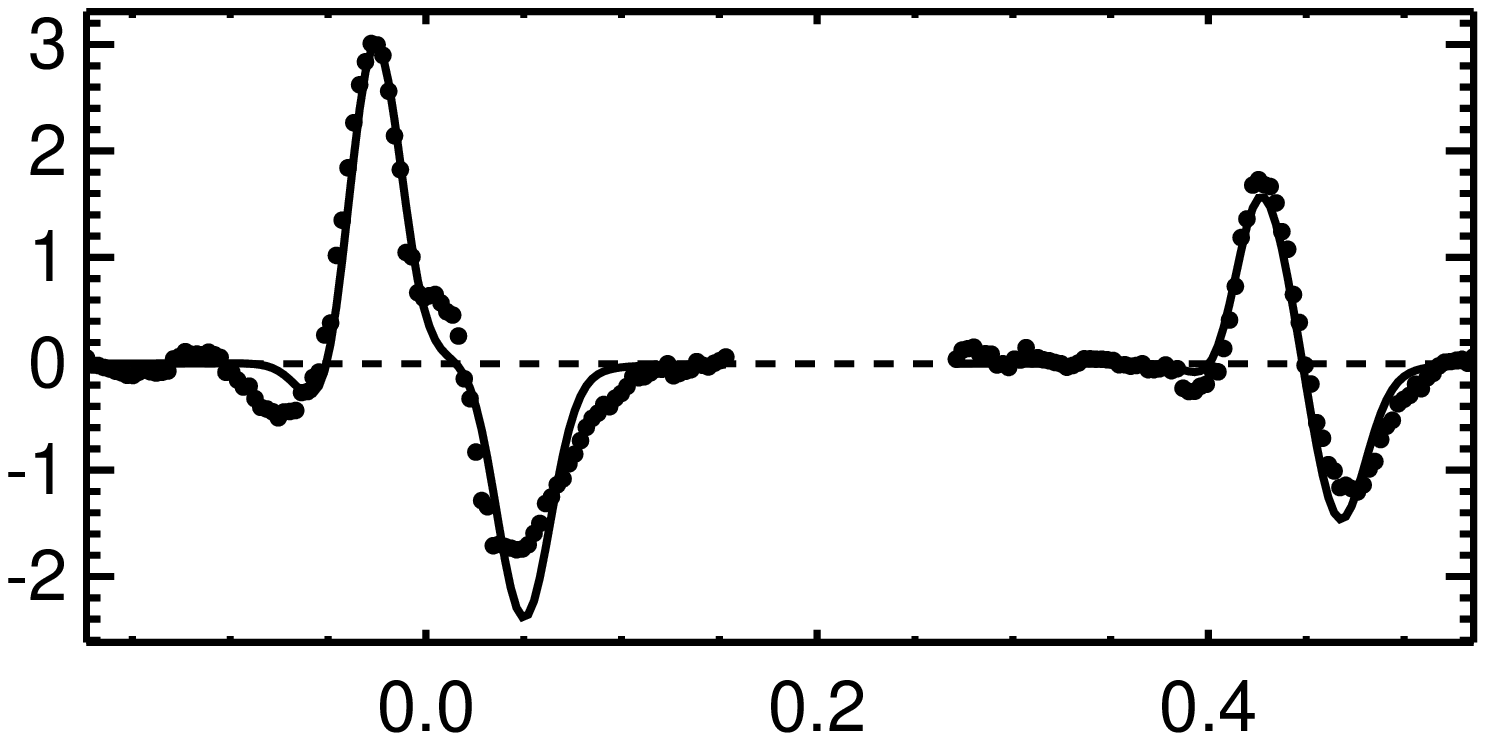}
\includegraphics[height=2.55cm,bb=70 25 473 273]{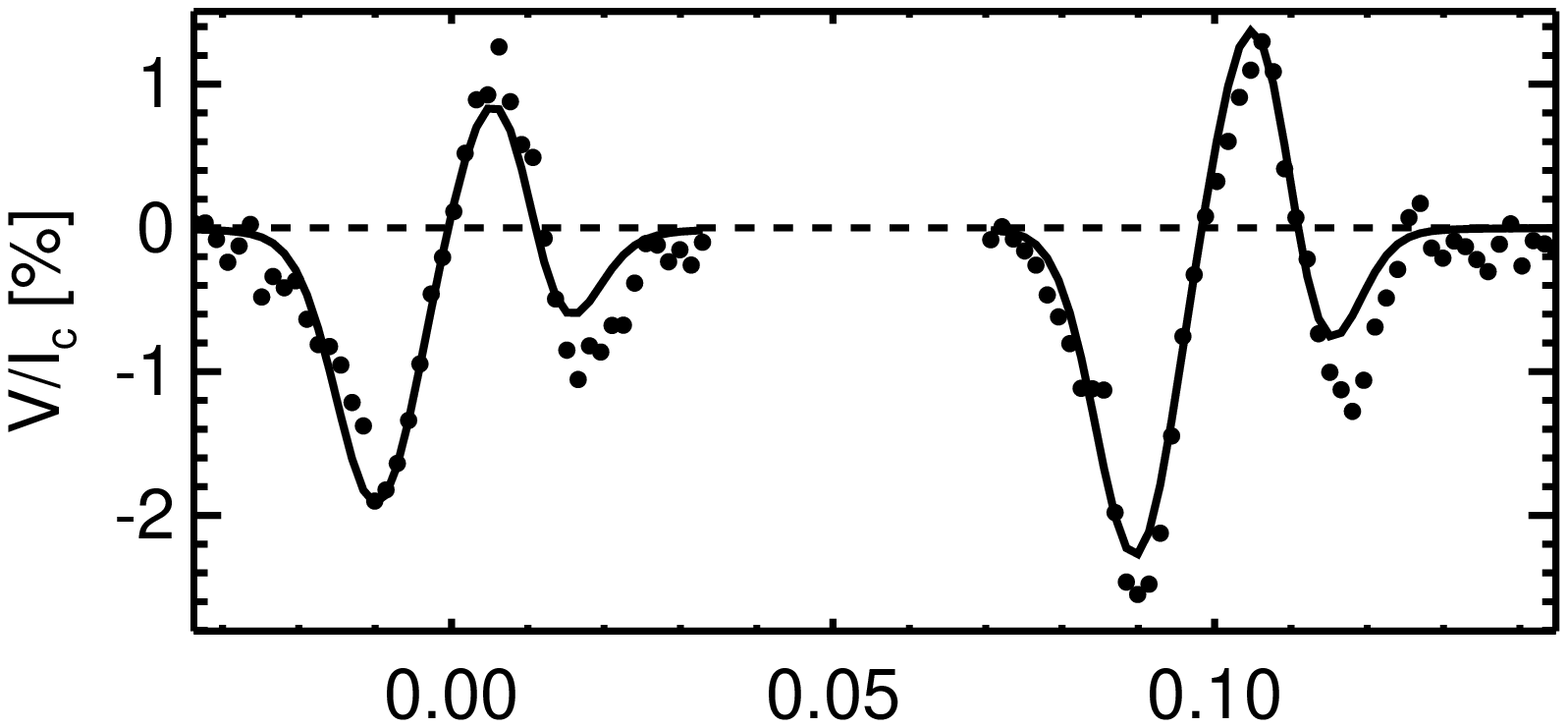}
\includegraphics[height=2.55cm,bb=30 25 508 273]{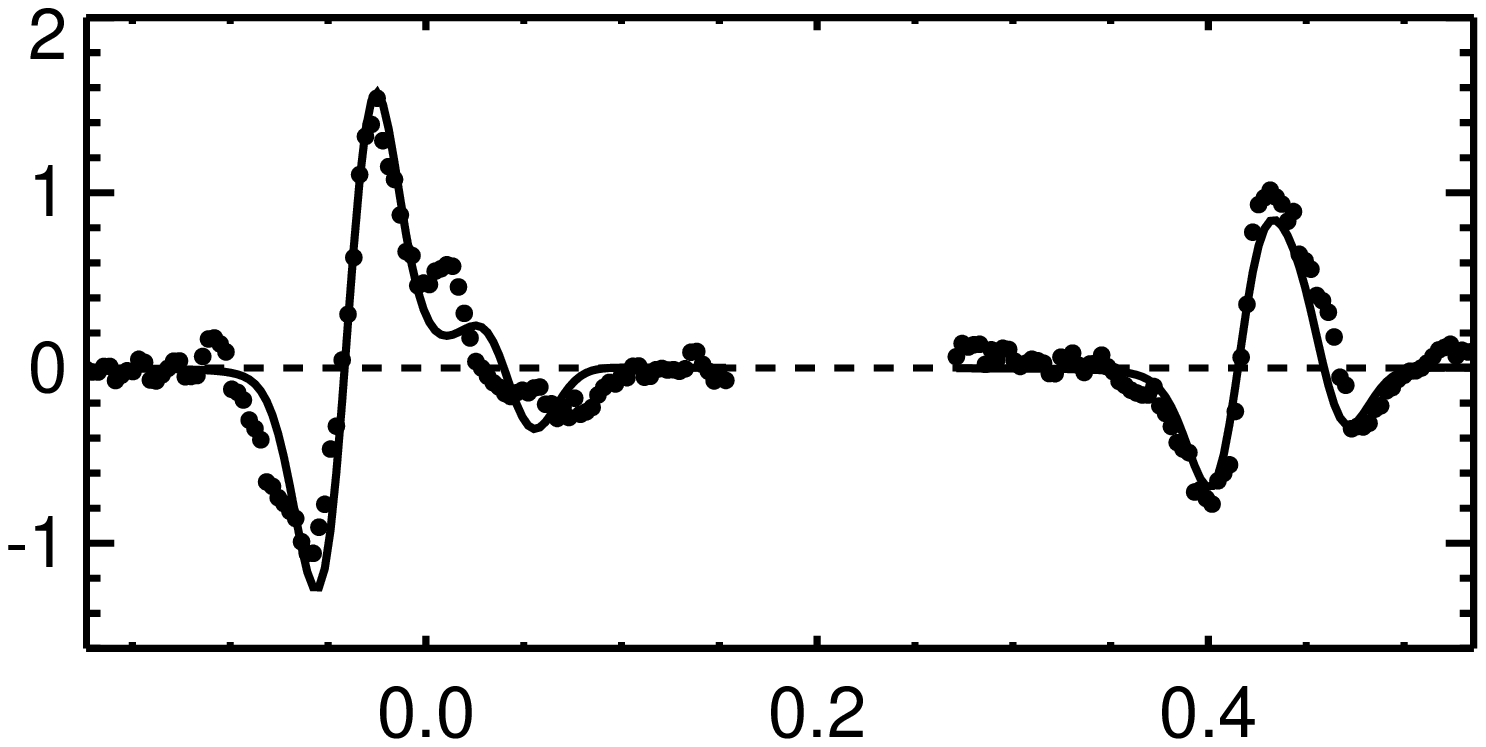}
\includegraphics[height=2.55cm,bb=70 25 473 273]{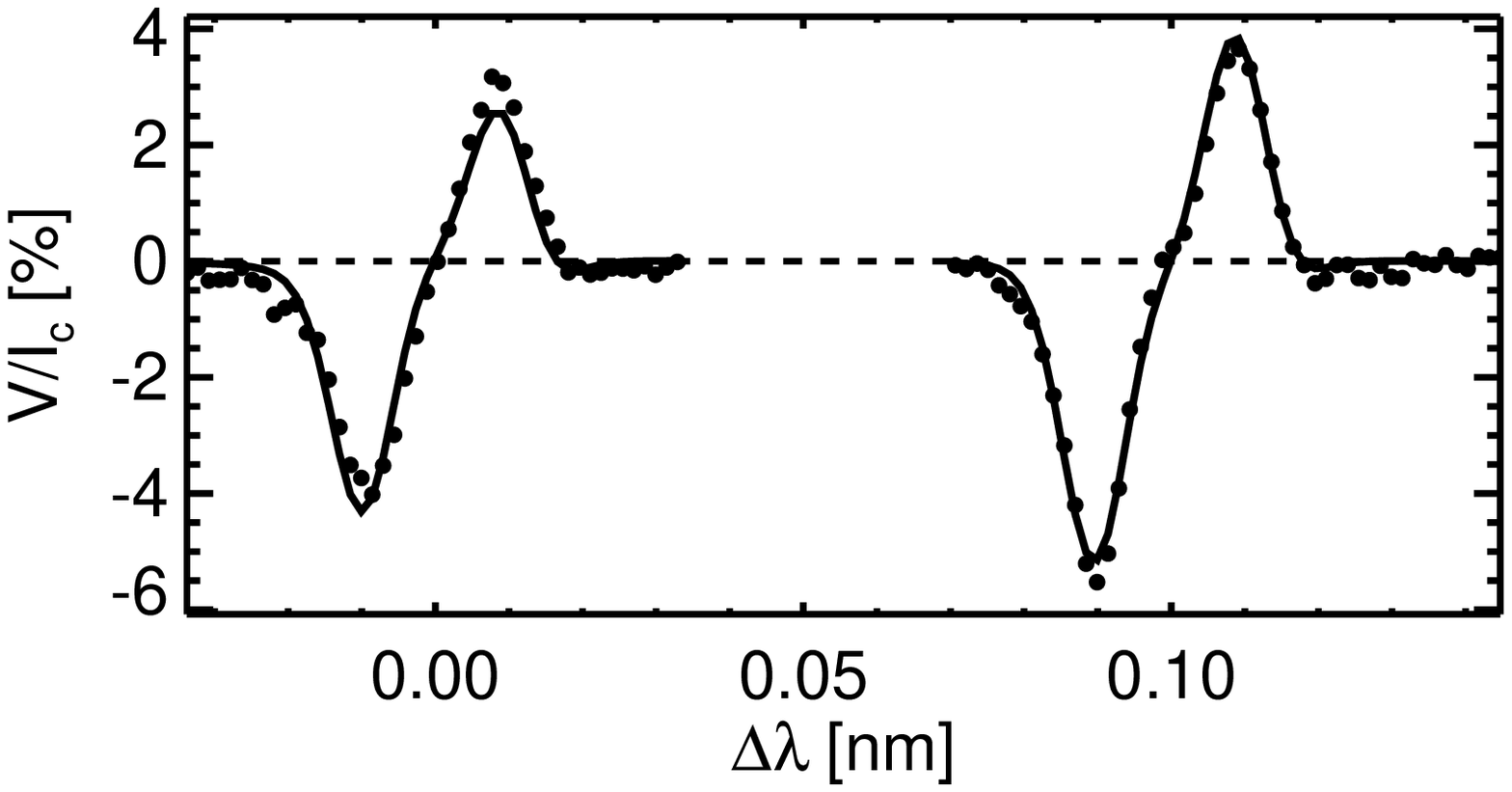}
\includegraphics[height=2.55cm,bb=30 25 508 273]{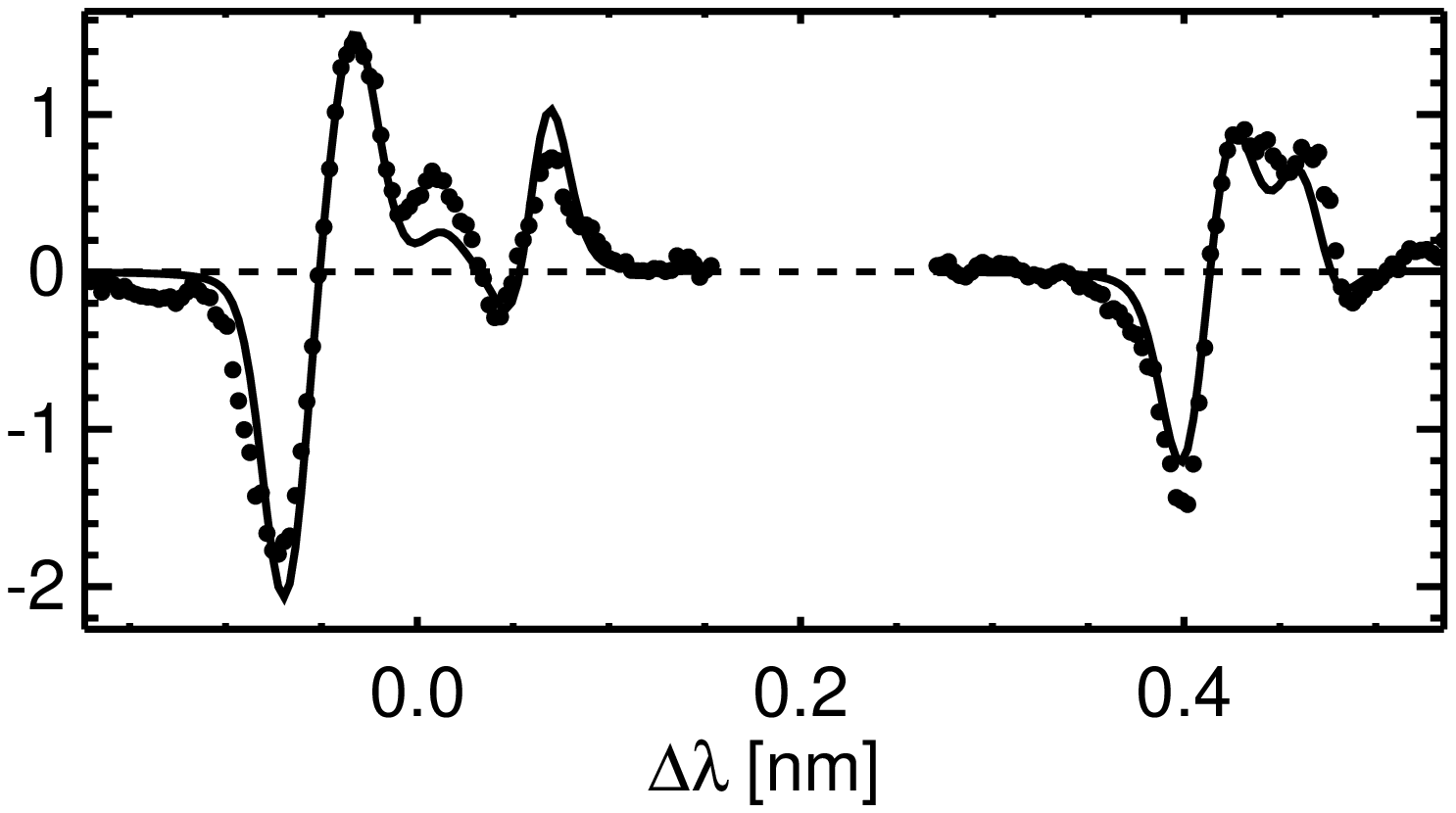}
\caption{Simultaneous spectropolarimetry of AR 10425 with TIP and POLIS at the
German VTT of Observatorio del Teide on August 9, 2003. {\em Left:} Stokes $V$
profiles of \ion{Fe}{i} 630.15 and 630.25~nm.  {\em Right:} Stokes $V$
profiles of \ion{Fe}{i} 1564.8 and 1565.2~nm.  The three rows correspond to
three different pixels in the limb-side penumbra ($\theta = 27^{\rm
o}$). Filled circles are the observations. Solid lines give the best-fit
profiles resulting from an uncombed inversion of the data using the code
described in \cite{bellotspw3}. Adapted from~\cite{beckthesis}.  }
\label{uncombedinversion}  
\end{figure}

As mentioned before, the uncombed model envisages the penumbra as a collection
of small magnetic flux tubes embedded in an ambient field. The thermal,
magnetic, and kinematic properties of the flux tubes and the ambient field
(Fig.~\ref{structure}) have been determined from Stokes inversions that use
two different magnetic atmospheres. These inversions \cite{bellotspw3,
bellotetal2004, borreroetal2004,borreroetal2005, beckthesis, borreroetal2006}
have demonstrated that the uncombed model is able to explain the shapes of the
polarization profiles of visible and infrared lines emerging from the penumbra
at resolutions of $\sim$1\arcsec\/ (see Fig.~\ref{uncombedinversion} for
examples). Perhaps the most important achievement of the model, however, is
that it quantitatively reproduces the NCP of visible
\cite{valentin,borreroetal2006} and infrared
\cite{borreroetal2005,muelleretal2006} lines, which are due to strong
gradients or discontinuities of the atmospheric parameters ({\em including
velocities}) along the line of sight. This success is not trivial, since the
spatial distribution of the NCP is determined primarily by discontinuities of
field inclination in the case of visible lines and discontinuities of field
azimuth in the case of infrared lines~\cite{landi,schlicheetal2002}.

As can be seen in Fig.~\ref{structure}, the tubes are inclined upward in the
inner penumbra and downward in the mid and outer penumbra.  The flow along the
tubes is parallel to the magnetic field at all radial distances. The agreement
is remarkable, but it hides a serious difficulty: a single flux tube cannot
extend across the penumbra with the inclinations of Fig.~\ref{structure},
because it would quickly leave the line forming region (even if the Wilson
depression is taken into account). A possible way out of this problem is that
the values shown in Fig.~\ref{structure} do not represent individual tubes,
but rather azimuthal averages over short flux tubes whose number density is
constant with radial distance (cf.\ \cite{wolfgang}).

The flux-tube properties and their radial variation, as derived from Stokes
inversions, agree well with those resulting from simulations of moving tubes 
in the thin tube approximation~\cite{schliche}. The simulations provide a
natural explanation for the Evershed flow in terms of a pressure gradient that
builds up along the tube as it rises buoyantly from the magnetopause and cools
off by radiative losses near the solar surface. The moving tube model explains
the motion of bright penumbral grains toward the umbra and the overall
morphology of penumbral filaments in continuum images. It also gives
convincing arguments why the flux tubes possess more horizontal and weaker
fields than the ambient atmosphere, and why the flux tubes return to the solar
surface in the mid and outer penumbra (i.e., why their field inclinations are
larger than 90$^{\rm o}$, cf.\ Fig.~\ref{structure}).  The apparent inability
of moving tubes to explain the surplus brightness of the penumbra
\cite{schliche+solanki} has been used by \cite{spruit+scharmer} as an argument
to propose the gappy penumbral model.  However, the remark made
in~\cite{schliche+solanki} that dissipation of the kinetic energy of the
Evershed flow could account for the penumbral brightness has been overlooked
by \cite{spruit+scharmer}.  Rejecting the idea of hot Evershed flows as the
origin of the penumbral brightness cannot be done without 2D or 3D simulations
of the evolution of flux tubes including a realistic energy equation and
stratified atmospheres.

The very existence of flux tubes embedded in a more vertical field has
been put into question alleging that such a configuration is not
force-free \cite{spruit+scharmer}. The imbalance of forces at the top
and bottom of the tubes would cause a vertical stretching that would
eventually destroy the tubes.  However, it has been demonstrated
\cite{borreroetal2007a} that the vertical stretching is limited by
buoyancy in convectively stable (subadiabatic) layers.  Also, it has
been shown that penumbral tubes can be brought into exact force
balance if the field within the tube has a small transversal component
\cite{borreroetal2007b}.  Interestingly, the temperature distributions
derived from the condition of magnetohydrostatic equilibrium of
penumbral tubes produce dark-cored filaments whose properties are very
similar to the observed ones \cite{borreroetal2007b}. The ability of
the uncombed model to explain the existence of dark-cored penumbral
filaments has also been demonstrated by means of 2D heat transfer
simulations of flux tubes carrying a hot Evershed flow
\cite{ruizcobo+bellot}.

From a modeling point of view, even the most complex Stokes inversions of
penumbral spectra use only two rays to describe the flux tube and the ambient
field, which is a very simplistic approximation (see \cite{bellotspw4} for
details).  Actually, the two rays represent homogeneous tubes with square
cross sections and ambient field lines that do not wrap around the tubes. More
sophisticated treatments of the uncombed penumbra are thus desirable for a
better interpretation of the observations. Such treatments could remove the
small differences between observed and best-fit profiles
(Fig.~\ref{uncombedinversion}).  However, one should not expect qualitatively
different results, since the uncombed models implemented in current inversion
codes already capture the essential physics needed to explain the shapes of
visible and infrared lines.

\subsection{MISMA penumbral model}

The MISMA model assumes that the penumbra is formed by optically thin magnetic
fibrils a few km in diameter \cite{jorge98,jorge}. Each resolution element
contains a messy bunch of field lines with random strengths and inclinations
that, for an unknown reason, are more or less parallel to the radial
direction. The model, implemented in practice as a simple two-component
atmosphere, successfully reproduces the asymmetries and NCPs of the
\ion{Fe}{i}~630.15 and 630.25~nm lines observed in sunspots at a 
resolution of $\sim$1\arcsec\/ \cite{jorge}.

According to MISMA inversions, downward flows with velocities that often
exceed 20~km~s$^{-1}$ exist everywhere in the penumbra \cite{jorge}. This
result is at odds with observations: 0\farcs2 resolution Dopplergrams show no
evidence for downflows in the inner and mid penumbra \cite{langhansetal2005}.
In addition, the mechanism whereby the small-scale fibrils get organized to
produce the large-scale (filamentary) structure of the penumbra remains
unknown. This is indeed a serious problem, because negligible azimuthal
fluctuations of magnetic field and velocity should be observed when both the
number of fibrils per resolution element is large and the fibrils follow the
same (random) distribution in different pixels.


As a proof of physical consistency, the MISMA deduced from the inversion was 
shown to satisfy the $\nabla \cdot \vec{B} =0$ condition, unlike 
simpler one-component models. However, azimuthally averaged atmospheric 
parameters were used rather than individual values. Since $\nabla \cdot \vec{B} 
=0$ must be verified {\em locally} pixel by pixel, this test does not 
really demonstrate the validity of the model.

It remains to be seen whether MISMAs are able to explain the shapes and NCPs
of infrared lines, as well as the existence of dark-cored penumbral
filaments. It is also necessary to find reasons why the magnetic fibrils that
form the lateral brightenings of dark-cored filaments know of each other so
well as to make them move coherently. If MISMAs are the building blocks of the
penumbra, regions with zero NCPs will not be detected even at high spatial
resolution, because there will always be fibrils interlaced along the
LOS. This is perhaps the most important prediction of the MISMA model.

\subsection{Gappy penumbral model}
\label{gappy}
The gappy model represents a theoretical attempt to explain the existence of
dark-cored penumbral filaments and the brightness of the penumbra
\cite{spruit+scharmer, scharmer+spruit}. It postulates that dark-cored
filaments are the signatures of radially oriented, field-free gaps located
just below the visible surface of the penumbra. Such gaps would sustain normal
convection, thereby providing energy to heat the penumbra.  This raises a 
serious problem, because the existence of vigorous field-free convection
plumes reaching the solar surface contradicts the accepted view
\cite{solanki+schmidt} that the penumbra is deep (as opposed to shallow).

Another problem is that it is not clear how the model can generate magnetic
fields pointing downward in the outer penumbra: the maximum field inclination
in a gappy penumbra is 90$^{\rm o}$, representing horizontal fields. Last, but
not least, the model does not offer any explanation for the Evershed flow. It
does not even have a suitable place to accommodate horizontal flows, because
they must reside where the field is nearly horizontal. Since this happens only
in very small volumes just above the gaps, a large fraction of the line
forming region would be devoid of flows.

The gappy model may be regarded as a limiting case of the uncombed model with
zero field strengths in the flux-tube component. The essential difference is
that a strong Evershed flow moves along the tube in the uncombed model,
whereas in a gappy penumbra not even the field-free regions harbor radial
outflows. Thus, an important ingredient for spectral line formation is missing
in the model: the discontinuous velocity stratifications produced by confined
Evershed motions {\em several} km~s$^{-1}$ in magnitude. Gappy models with
potential fields do exhibit gradients of field strength, inclination, and
azimuth with height \cite{scharmer+spruit}, but it is unlikely that such
gradients can reproduce the {\em multi-lobed} Stokes $V$ profiles and the NCPs
of spectral lines without including strong Doppler shifts in an ad hoc
manner. Convection in the field-free gaps alone will not produce
large NCPs or multi-lobed profiles because (a) it occurs near $\tau =1$, i.e.,
far from the line forming region, and (b) the associated velocities will
certainly be smaller than 5-6~km~s$^{-1}$.

In summary, although the idea may be appealing, radiative transfer
calculations must be performed to demonstrate that the gappy model is
able to reproduce the spectropolarimetric properties of the
penumbra. Also, heat transfer simulations are required to prove that
the field-free gaps would indeed be observed as dark-cored filaments,
and that the gaps can heat the penumbra to the required degree.
Without these calculations, it seems premature to
accept the gappy model as a good representation of sunspot penumbrae.

\section{Outlook}
Currently available models of the penumbra have both strengths and
limitations. The difference is that the uncombed model has been extensively
confronted with observations, while the MISMA and gappy models still need to
pass stringent observational tests to demonstrate their plausibility. Some of
the basic claims made by the later models have not yet been confirmed by
radiative and/or heat transfer calculations, and hence remain speculative.

Further advances in our understanding of the penumbra will come from
spectropolarimetric observations at 0\farcs2--0\farcs3. This is the minimum
resolution needed to identify the dark cores of penumbral filaments.  We would
like to measure the vector magnetic fields and velocities of dark-cored
filaments not only to distinguish between competing models (which imply
different convection modes in the presence of inclined fields), but also to
drive holistic MHD simulations of the penumbra.  The required observations
will be obtained with instruments like the Spectro-Polarimeter~\cite{hinode}
aboard HINODE, TIP~\cite{collados} at GREGOR, IMaX \cite{imax} onboard
SUNRISE, and VIM \cite{vim} aboard Solar Orbiter.


\printindex
\end{document}